\documentclass[runningheads]{llncs}

\usepackage{graphicx}
\usepackage[utf8]{inputenc}
\usepackage{amssymb}
\usepackage{hyperref}
\usepackage{mathtools}
\usepackage{amsmath}
\usepackage{enumerate}
\usepackage{tikz}
\usepackage{amsmath}
\usepackage{amssymb}
\usepackage{algorithm}
\usepackage[noend]{algpseudocode}
\usepackage{multirow}
\usepackage{enumitem}
\usepackage{array}
\usepackage{comment}
\usepackage{subcaption}
\usepackage{caption}

\begin{document}
\title{M-VAAL: Multimodal Variational Adversarial Active Learning for Downstream Medical Image Analysis Tasks}

\authorrunning{B. Khanal et al.}

\author{Bidur Khanal\inst{1}\and
Binod Bhattarai\inst{4}\and
Bishesh Khanal\inst{3}\and
Danail Stoyanov \inst{5} \and
Cristian A. Linte\inst{1,2}}


\titlerunning{Multimodal Variational Adversarial Active Learning}
%

\institute{ Center for Imaging Science, RIT, Rochester, NY, USA \and Biomedical Engineering, RIT, Rochester, NY, USA \and NepAl Applied Mathematics and Informatics Institute for Research (NAAMII) \and University of Aberdeen, Aberdeen, UK\\ \and University College London, London, UK}

\maketitle              
\begin{abstract}
Acquiring properly annotated data is expensive in the medical field as it requires experts, time-consuming protocols, and rigorous validation. Active learning attempts to minimize the need for large annotated samples by actively sampling the most informative examples for annotation. These examples contribute significantly to improving the performance of supervised machine learning models, and thus, active learning can play an essential role in selecting the most appropriate information in deep learning-based diagnosis, clinical assessments, and treatment planning. Although some existing works have proposed methods for sampling the best examples for annotation in medical image analysis, they are not task-agnostic and do not use multimodal auxiliary information in the sampler, which has the potential to increase robustness. Therefore, in this work, we propose a Multimodal Variational Adversarial Active Learning (M-VAAL) method that uses auxiliary information from additional modalities to enhance the active sampling. We applied our method to two datasets: i) brain tumor segmentation and multi-label classification using the BraTS2018 dataset, and ii) chest X-ray image classification using the COVID-QU-Ex dataset. Our results show a promising direction toward data-efficient learning under limited annotations.

\end{abstract}

\section{Introduction}



Automated medical image analysis tasks, such as feature segmentation and disease classification play an important role in assisting with clinical diagnosis, as well as appropriate planning of non-invasive therapies, including surgical interventions \cite{hamamci2011tumor,ansari2022practical}.
In recent years, supervised deep learning-based methods have shown promising results in clinical settings. However, clinical translation of supervised deep learning-based models is still limited for most applications, due to the lack of access to a large pool of annotated data and generalization capability.

As medical data is expensive to annotate, some methods have explored the generation of synthetic images with ground truth annotation \cite{singh2021medical}.
However, generative models for synthesizing high-quality medical data have not yet achieved a state where their distribution perfectly matches the distribution of real medical data, especially those containing rare cases \cite{skandarani2023gans,al2023usability}. This limitation can often result in biased models and poor performance. Another approach with a growing interest is semi-supervised learning where very few labeled data along with a large number of unlabeled data is used to train deep learning models \cite{luo2021dtc,chen2021semi,verma2019interpolation}.
Nevertheless, in a real-world scenario, semi-supervised learning still requires the selection of a certain number of image samples to be annotated by experts.

Active learning attempts to sample the best subset of examples for annotation from a pool of unlabeled examples to maximize the downstream task performance.
Methods to sample examples that provide the best improvement with a limited budget have a long history \cite{lewis1995sequential}, and several approaches have been proposed in recent years \cite{zhan2022comparative}.
Active learning (AL) has experienced increased traction in the medical imaging domain, as it enables a human-in-the-loop approach to improve deployed AI models \cite{budd2021survey}.
Recent works have proposed various sampling methods in AL specific to tasks such as classification, segmentation, and depth estimation in medical imaging.
Shao \textit{et al}. \cite{Shao} used active sampling to minimize annotation for nucleus classification in pathological images.
Yang \textit{et al}. \cite{yang2017suggestive} proposed a framework to improve lymph node segmentation from ultrasound images using only $50\%$ training data. Laradji \textit{et al}. \cite{laradji2020weakly} adopted an entropy-based method for a region-based active sampler for COVID-19 lesion segmentation from CT images.
Thapa \textit{et al}. \cite{thapa2022task} proposed a task-aware AL for depth estimation and segmentation from endoscopic images. Related to our work, Sharma \textit{et al}. \cite{sharma2019active} used uncertainty and representativeness to actively sample training examples for 2D brain tumor segmentation. Lastly, Kim \textit{et al}. \cite{kim2023active} studied various AL approaches for 3D brain tumor segmentation task.

\textit{Task Agnostic:} The methods described earlier require training of downstream task-specific models precluding the possibility of training a task-agnostic model from a given pool of unannotated input images when the intended downstream task is not known a priori.
On the other hand, task-agnostic methods could enable the use of the same model for new tasks that may arise in real-world continuous deployment settings.
VAAL \cite{sinha2019variational} is a promising task-agnostic method that trains a variational auto-encoder (VAE) with adversarial learning.
VAAL produces a low-dimensional latent space that aims to capture both uncertainty and representativeness from the input data, enabling effective sampling for AL.

\textit{Sampling from multimodal images:} Clinicians rarely reach a clinical decision by looking into a single medical image of a patient.
They rely on other information, such as clinical symptoms, patient reports, multimodal images, and auxiliary device information.
We believe that sampling methods in AL could benefit from using multimodal information.
Although some recent works have explored multimodal medical images in the context of AL \cite{budd2021survey}, none of the existing methods directly use the multimodal image as auxiliary information to learn the sampler.

In this paper, we propose a task-agnostic method that exploits multimodal imaging information to improve the AL sampling process. We modify the existing task-agnostic VAAL framework to enable exploiting multimodal images and evaluate its performance on two widely used publicly available datasets: the multimodal BraTS dataset \cite{brats} and COVID-QU-Ex dataset \cite{anas_2022} containing lung segmentation maps as additional information.

The contributions of this work are as follows: \textbf{1)} We propose a novel multimodal variational adversarial AL method (M-VAAL) that uses additional information from auxiliary modalities to select informative and diverse samples for annotation; \textbf{2)} Using the BraTS2018 and COVID-QU-Ex datasets, we show the effectiveness of our method in actively selecting samples for segmentation, multi-label classification, and multi-class classification tasks. \textbf{3)} We show that the sample selection in AL can potentially benefit from the use of auxiliary information. 


\section{Methodology}
\begin{figure}[h!]
\begin{center}
\includegraphics[width=1\columnwidth]{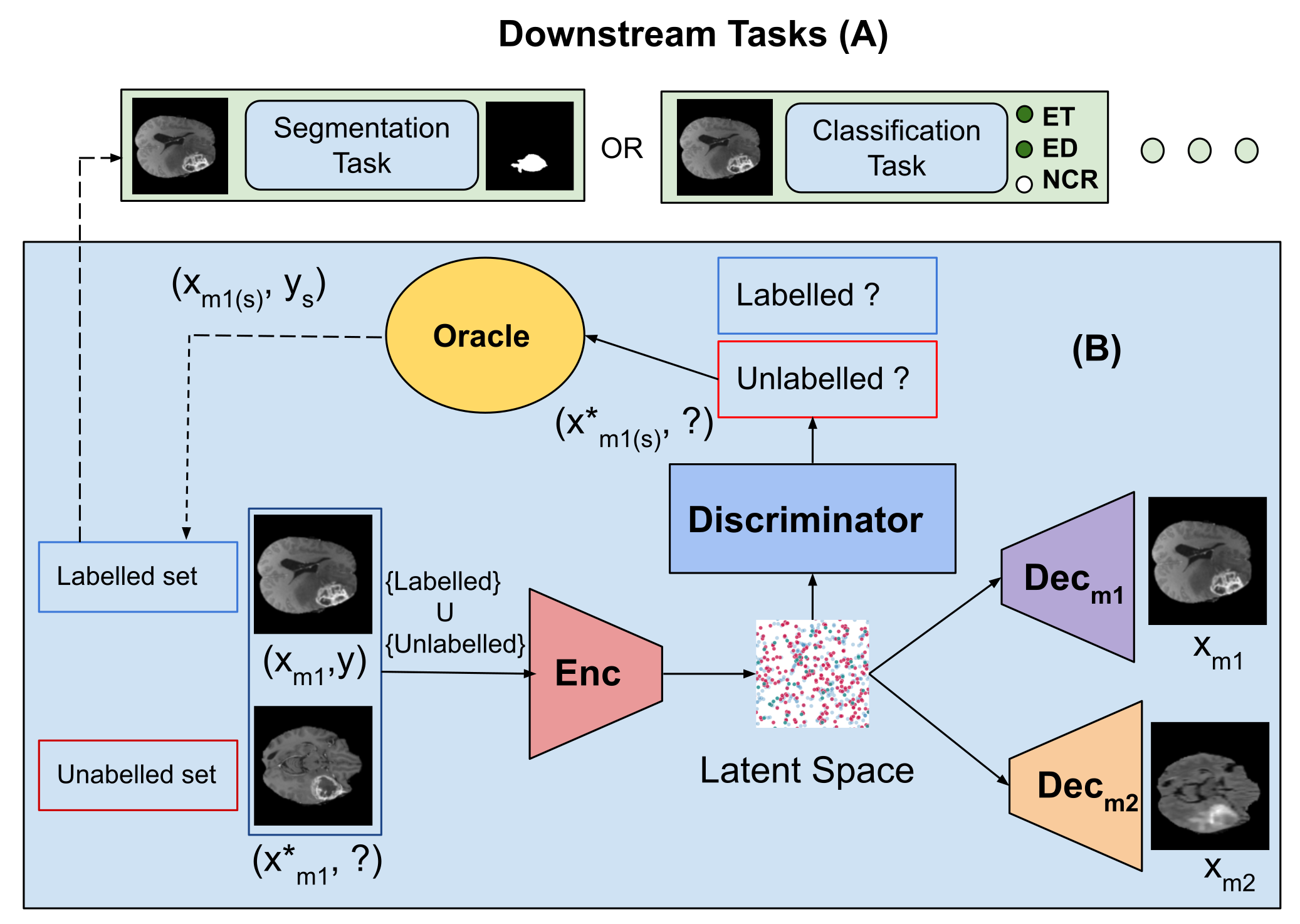}
\caption{{M-VAAL Pipeline: Our active learning method uses multimodal information ($m1$ and $m2$) to improve VAAL. M-VAAL samples the unlabelled images, and selects samples complementary to the already annotated data, which are passed to Oracle for annotation. Incorporating auxiliary information from the second modality ($m2$) produces a more generalized latent representation for sampling. Our method learns task-agnostic representations, therefore the latent space can be used for both classification and segmentation tasks to sample the best-unlabelled images for annotation. (Refer to \ref{m-vaal-sampler} for the meaning of each notation)
\label{fig:overall_method}%
}}
\end{center}
\end{figure}

In active learning, a subset of most informative samples $X_{s}$ is selected from a large pool of unlabelled set $X^{*}$ to query labels $Y_{s}$. The number of samples selected for label annotation is dictated by the set budget $b$ of the sampler. Let us consider $(x,y) \sim (X,Y)$ as a labeled data pair initially present in the dataset. After sampling $b$ examples from an unlabelled pool $X^{*}$, they are labeled and added to the labeled pool $(X,Y)$. The sampling process is iterative, such that $b$ examples are queried at each active sampling round and added to the labeled pool. The labeled samples are used to train a task network at each round by minimizing the task objective function. In our study, we have considered three different downstream tasks: a) semantic segmentation of brain tumors, b) multi-label classification of tumor types, and c) multi-class classification of chest conditions. The overall workflow of our proposed method is shown in Fig. \ref{fig:overall_method}.

\subsection{Task Learner (A)}

Our objective is to enhance the performance of downstream tasks by utilizing the smallest number of labeled samples. The task network is defined by the parameter $\theta_T$. For the segmentation task, we train a U-Net model \cite{ronneberger2015u} to generate pixel-wise segmentation maps $y'$, given a labeled sample $(x,y) \sim (X,Y)$, where $y$ is a binary image. We chose the U-Net model due to its simplicity, as our focus is on evaluating the effectiveness of AL, not on developing a state-of-the-art segmentation technique.

For the classification task, we use a ResNet-18 \cite{he2016identity} classifier to predict $y'$, given $(x,y) \sim (X,Y)$ where $y \subseteq \{0,1,2,..\}$. In a multi-label classification setting, $y'$ can contain more than one class, while in a multi-class setting, $y'$ belongs to only one class from the set. Once a task is trained using labeled samples, we move on to the sampler (B) stage to select the best samples for annotation. After annotating the selected samples, we add them to the labeled sample pool and retrain the task learner using the updated sample set. Thus, the task learner is completely retrained for multiple rounds, with an increased training budget each time.
\subsection{M-VAAL Sampler (B)}
\label{m-vaal-sampler}
Our proposed M-VAAL method extends the VAAL approach by using an encoder-decoder architecture combined with adversarial learning to generate a lower dimensional latent space that captures sample uncertainty and representativeness. Unlike other AL methods that directly estimate uncertainty or diversity using task representation, VAAL learns a separate task-agnostic representation, providing the freedom to model uncertainty independent of task representation. To further strengthen the representation learning capability, we introduce a second modality as auxiliary information. We modify the VAE to reconstruct the original input image from modality $m1$ and the auxiliary image of modality $m2$ using the same latent representation generated from the $m1$ images. The objective function of the VAE is formulated as:
\begin{equation}
\small
\label{m1_decoder}
\begin{aligned}
\mathcal{L}_{\mathrm{VAE}}^{\text{\textit{m1}}}=& \mathbb{E}\left[\log p_{\theta_{m1}}\left(x_{m1} \mid z\right)\right] +\mathbb{E}\left[\log p_{\theta_{m1}}\left(x^{*}_{m1} \mid z^{*}\right)\right]
\end{aligned}
\end{equation}
\begin{equation}
\small
\label{m2_decoder}
\begin{aligned}
\mathcal{L}_{\mathrm{VAE}}^{\text{\textit{m2}}}=& \mathbb{E}\left[\log p_{\theta_{m2}}\left(x_{m2} \mid z\right)\right] +\mathbb{E}\left[\log p_{\theta_{m2}}\left(x^{*}_{m2} \mid z^{*}\right)\right]
\end{aligned}
\end{equation}
where $p_{\theta_{m1}}$ and $p_{\theta_{m2}}$ are the decoders for modality $m1$ and $m2$, respectively, parameterized by $\theta_{m1}$ and $\theta_{m2}$. Likewise, $z$, $x_{m1}$, and $x_{m2}$ represent a latent representation, an image of modality $m1$, and an image of modality $m2$, respectively, of the samples belonging to the labelled set; similarly, $z^{*}$, $x^{*}_{m1}$, and $x^{*}_{m2}$ represent a latent representation, an  image of modality $m1$, and an image of modality $m2$, respectively, of the samples belonging to the unlabelled set. The VAE also uses a Kullback-Leibler (KL) distance loss to regularize the latent representation, formulated as:
\begin{equation}
\small
\begin{aligned}
\mathcal{L}_{\mathrm{VAE}}^{KL}=&- \mathrm{D}_{\mathrm{KL}}\left(q_{\phi_{m1}}\left(z \mid x_{m1}\right)|| p(z)\right) - \mathrm{D}_{K L}\left(q_{\phi_{m1}}\left(z^{*} \mid x^{*}_{m1}\right)|| p(z)\right)
\end{aligned}
\end{equation}

where $q_{\phi_{m1}}$ is the encoder of modality $m1$ parameterized by $\phi_{m1}$ and $p(z)$ is the prior chosen as a unit Gaussian. It should be noted that only a single encoder is used to generate the latent representation, while there are two decoders.
\begin{figure}[h!]
\begin{center}
\includegraphics[width=0.7\columnwidth]{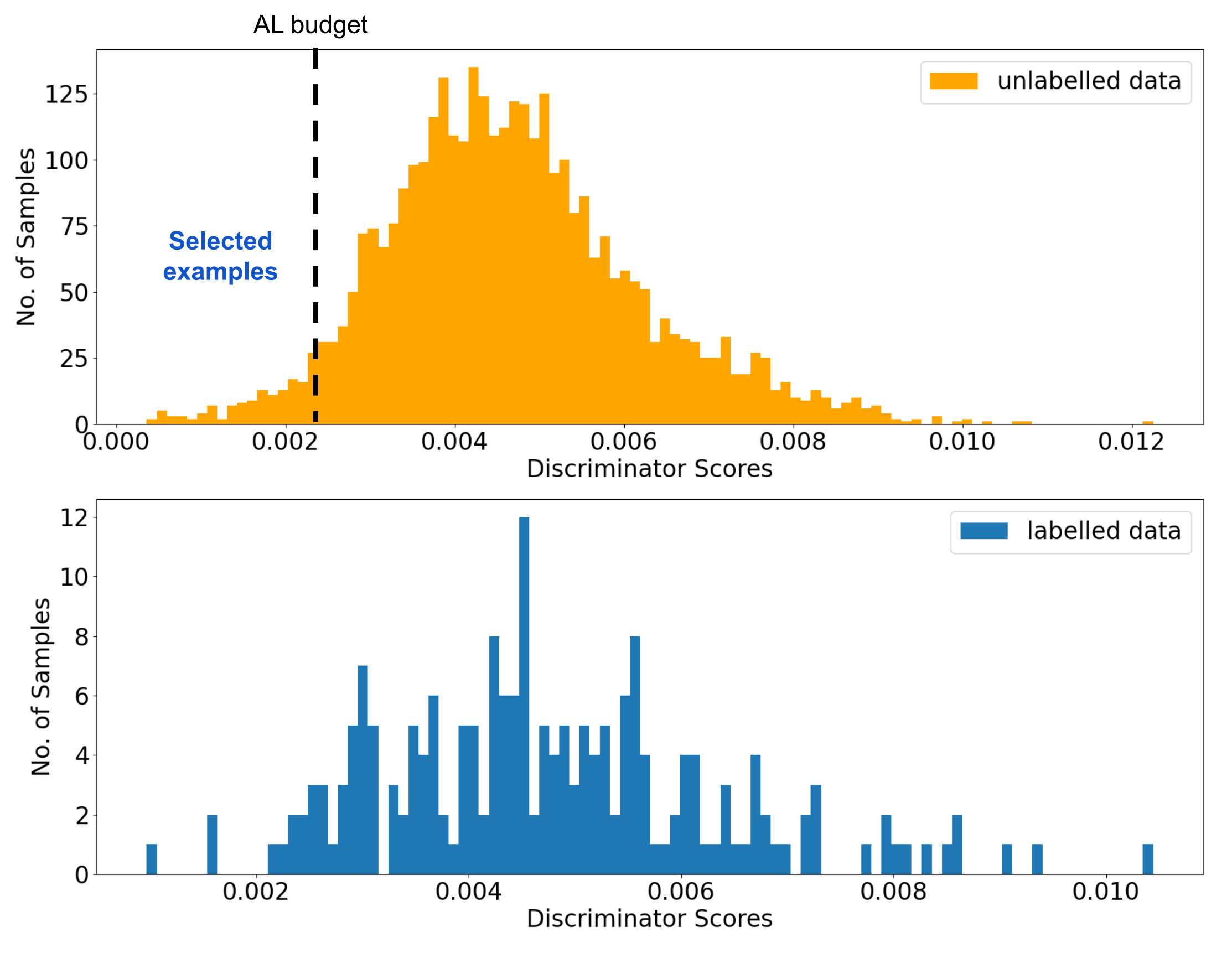}
\caption{{Histogram of the discriminator scores for unlabeled and labeled data at third AL round. The discriminator is adversarially trained to push unlabeled samples toward lower values and labeled samples toward higher values. Our method involves selecting unlabelled instances that are far from the peak distribution of the labeled data. The number of samples to select is dictated by the AL budget.
\label{fig:decision_plot}%
}}
\end{center}
\end{figure}
Finally, an adversarial loss is added to encourage the VAE towards generating the latent representation that can fool the discriminator in distinguishing labeled from unlabelled examples (shown in Eq. \ref{vae_adv}). We train both the VAE and the discriminator in an adversarial fashion.
The discriminator (D) is trained on the latent representation of an image to distinguish if an image comes from a labeled set or an unlabeled set. The loss function for the discriminator is shown in Eq. \ref{disc_eqn}. We used Wasserstein GAN loss with gradient penalty \cite{gulrajani2017improved}. 

\begin{equation}
\small
\label{vae_adv}
    \mathcal{L}_{\mathrm{VAE}}^{a d v}=\mathbb{E}\left[D\left(q_{\phi_{m1}}\left(z \mid x_{m1}\right)\right)\right]-\mathbb{E}\left[D\left(q_{\phi_{m1}}\left(z^{*} \mid x^{*}_{m1}\right)\right)\right]
\end{equation}
\begin{equation}
\small
\label{disc_eqn}
    \mathcal{L}_D=-\mathbb{E}\left[D\left(q_{\phi_{m1}}\left(z \mid x_{m1}\right)\right)\right]+\mathbb{E}\left[D\left(q_{\phi_{m1}}\left(z^{*} \mid x^{*}_{m1}\right)\right)\right] + \lambda \mathbb{E} \left [\left(\left\|\nabla_{\hat{x}} D(\hat{x})\right\|-1\right)^2\right]
\end{equation}
where the third term in Eq. \ref{disc_eqn} is the gradient penalty and $\hat{x}=\epsilon x_{m1}+(1-\epsilon) x^{*}_{m1}$ is a randomly weighted average between $x_{m1}$ and $x^{*}_{m1}$, such that $0 \le \epsilon \le 1$.

\begin{algorithm}[ht!]
    \small
    \caption{Multimodal Variational Adversarial Active Learning} \label{alg:vaal}
    \begin{algorithmic}[1]
        \renewcommand{\algorithmicrequire}{\textbf{Given:}}
        \Require Hyperparameters: epochs, $\gamma_1$, $\gamma_2$, $\gamma_3$, $\delta_1$,  $\delta_2$,  $\delta_3$
        \renewcommand{\algorithmicrequire}{\textbf{Input:}}
        \Require Labeled data ($x_{m1}, y, x_{m2})$, Unlabeled data $(x^{*}_{m1}, x^{*}_{m2})$
        \renewcommand{\algorithmicrequire}{\textbf{Initialize:}}
        \renewcommand{\algorithmicensure}{\textbf{Initialize:}}
        \Require Model parameters $\theta_{T}$, $\theta_{VAE} = \{ \theta_{m1}$,$\theta_{m2}$,$\phi$\}, and $\theta_{D}$
        
        \For {$e = 1$ \text{to epochs}}
        \State sample $(x_{m1}, y, x_{m2}) \sim (X_{m1}, Y, X_{m2}$)
        \State sample $(x^{*}_{m1}, x^{*}_{m2}) \sim (X^{*}_{m1},X^{*}_{m2}$)
        \vspace{3pt}     
        \State $\mathcal{L}_{\mathrm{VAE}} \gets \gamma_1 \mathcal{L}_\mathrm{VAE}^{adv} + \gamma_2 \mathcal{L}_{\mathrm{VAE}}^{m1} +  \gamma_3 \mathcal{L}_\mathrm{VAE}^{m2} + \mathcal{L}_\mathrm{VAE}^{KL}$ 
        \vspace{1pt}
        \State Update VAE: $\theta'_{VAE} \gets \theta_{VAE} - \delta_1 \nabla \mathcal{L}_{\mathrm{VAE}} $
        \vspace{1pt}
        \State Update $D$: $\theta'_{D} \gets \theta_{D} - \delta_2 \nabla \mathcal{L}_\mathrm{D} $
        \State Train and update $T$: $\theta'_{T} \gets \theta_T - \delta_3 \nabla \mathcal{L}_{\mathrm{T}} $
        \EndFor
    \end{algorithmic}
    \vspace{-0.3cm}
    \hrulefill
    \begin{algorithmic}[1]
        \vspace{-0.1cm}
        \renewcommand{\algorithmicrequire}{\textbf{\textit{Sampling Phase}}}
        \Require 
        \renewcommand{\algorithmicrequire}{\textbf{Input:}}
        \renewcommand{\algorithmicensure}{\textbf{Output:}}
        \Require $b, X_{m1}, X^{*}_{m1}$
        \Ensure $X_{m1}, X^{*}_{m1}$
        \State Select samples $X^{*}_{m1(s)}$ with $\min_b\{\theta_{D}(z^{*})\}$ 
        \State $y_{s} \leftarrow \mathcal{ORACLE}(X^{*}_{m1(s)})$
        \State $(X_{m1},Y) \leftarrow (X_{m1},Y) \cup (X^{*}_{m1(s)},Y_{s})$
        \State $X^{*}_{m1} \leftarrow X^{*}_{m1} - X^{*}_{m1(s)}$\\
        \Return $X_{m1}, X^{*}_{m1}$
    \end{algorithmic} 
\end{algorithm}

During the sampling process, the top samples that the discriminator votes as belonging to the unlabelled set are chosen; our intuition is that these samples contain information most different from the ones carried by the samples already belonging to the labeled set. Fig. \ref{fig:decision_plot} shows that our method selects the examples that are far from the distribution of labeled examples, thus capturing diverse samples.
The overall training and sampling sequence of M-VAAL is shown in Algorithm \ref{alg:vaal}.
In the first round, the task network is trained only with the initially labeled samples to minimize the task objective function $\mathcal{L}_{T}$.
Simultaneously, the M-VAAL sampler is also trained independently using all the labeled and unlabelled examples, with the end goal of improving the discriminator at distinguishing between labeled and unlabelled pairs.
After the initial round of training, the trained discriminator is used to select the top $b$ unlabelled samples identified as belonging to the unlabelled set.
These selected samples are sent to the Oracle for annotation. Finally, the selected samples are removed from the unlabelled set and then added to the pool of labeled sets, along with their respective labels.
The next round of AL proceeds in a similar fashion, but with the updated labeled and unlabelled sets.

\section {Experiments}
\subsection{Dataset}
\subsubsection{BraTS2018:}
We used the BraTS2018 dataset \cite{brats}, which includes co-registered 3D MR brain volumes acquired using different acquisition protocols, yielding slices of T1, T2, and T2-Flair images. To sample informative examples from unlabelled brain MR images, we employed the M-VAAL algorithm using contrast-enhanced T1 sequences as the main modality, and T2-Flair as auxiliary information. Contrast-enhanced T1 images are preferred for tumor detection as the tumor border is more visible \cite{bouget2021meningioma}. In addition, T2-Flair, which captures cerebral edema (fluid-filled region due to swelling), can also be utilized for diagnosis \cite{brats}. Our focus was on the provided 210 High-Grade Gliomas (HGG) cases, which included manual segmentation verified by experienced board-certified neuro-radiologists. There are three foreground classes: Enhancing Tumor (ET), Edematous Tissue (ED), and Non-enhancing Tumor Core (NCR). In practice, the given foreground classes can be merged to create different sub-regions such as whole tumor and tumor core for evaluations \cite{brats}. The whole tumor entails all foreground classes, while the tumor core only entails ET and NCR. 

Before extracting 2D slices from the provided 3D volumes, we randomly split the 210 volumes into training and test cases with an 80:20 ratio. The training set was further split into training and validation cases using the same ratio of 80:20, resulting in 135 training, 33 validation, and 42 test cases. These splits were created before extracting 2D slices to avoid any patient information leakage in the test splits. Each 3D volume had 155 ($240 \times 240$) transverse slices in axial view with a spacing of 1 mm. However, not all transverse slices contained tumor regions, so we extracted only those containing at least one of the foreground classes. Some slices contained only a few pixels of the foreground segmentation classes, so we ensured that each extracted slice had at least 1000 pixels representing the foreground class; any slice not meeting this threshold was discarded. Consequently, the curated dataset comprised 3673 training images, 1009 validation images, and 1164 test images of contrast-enhanced T1, T2-Flair, and the segmentation map.

We evaluated our method on two downstream tasks: whole tumor segmentation and multi-label classification task. In the multi-label classification task, our prediction classes consisted of ET, ED, and NCR, with each image having either one, two, or all three classes. It's worth noting that for the downstream task, only contrast-enhanced T1 images were used, while M-VAAL made use of both contrast-enhanced T1 and T2-Flair images.
 
\subsubsection{COVID-QU-Ex:} The COVID-QU-Ex dataset is composed of $256 \times 256$ chest X-ray images from different patients that have been categorized into one of three groups: COVID infection, Non-COVID infection, and Normal. The dataset has two subsets: lung segmentation data and COVID-19 infection segmentation data, and the latter was chosen for our experiment. In addition to the X-ray images, the dataset also provides a segmentation mask for each image, which was utilized as an auxiliary modality during training M-VAAL. The dataset has a total of 5,826 images, consisting of 1,456 Normal, 1,457 Non-COVID-19, and 2,913 COVID-19 cases. These images are split into three sets: training, validation, and test, with the training set containing 3,728 images, the validation set containing 932 images, and the test set containing 1,166 images. The downstream task was to classify the input X-ray image into one of three classes.

\subsection{Implementation Details}

\subsubsection{BraTS2018:}
All the input images have a single channel of size $240 \times 240$. To preprocess the images, we removed $1\%$ of the top and bottom intensities and normalized them linearly by dividing each pixel with the maximum intensity, bringing the pixel intensity to a range of 0 to 1. We also normalized the images by subtracting and dividing them by the mean and standard deviation, respectively, of the training data. For VAAL and M-VAAL, the images were center-cropped to 210 x 210 pixels and resized to $128 \times 128$ pixels. For the downstream task, we used the original image size. Furthermore, to stabilize the training of VAAL and M-VAAL under similar hyperparameters as the original VAAL \cite{sinha2019variational}, we converted the single-channel input image to a three-channel RGB with the same grayscale value across all channels.

For M-VAAL, we used the same $\beta$-VAE and discriminator as in original VAAL \cite{sinha2019variational}, but added a batch normalization layer after each linear layer in the discriminator. Additionally, instead of using vanilla GAN with binary-cross entropy loss, we used WGAN with a gradient penalty to stabilize the adversarial training \cite{gulrajani2017improved}, with $\lambda = 1$. The latent dimension of VAE was set to 64, and the initial learning rate for both the VAE and discriminator was $1 e^{-4}$. We tested $\gamma_3$ in the range $M = {0.2,0.4,0.8,1}$ via an ablation study reported in Sec \ref{ablation_study}, while $\gamma_1$ and $\gamma_2$ were set to 1. Both VAAL and M-VAAL used a mini-batch size of 16 and were trained for 100 epochs using the same hyperparameters, except for $\gamma_3$, which was only present in M-VAAL. For consistency, we initialized the model at each stage with the same random seed and repeated three trials with three different seeds for each experiment, recording the average score.

For the segmentation task, we used a U-Net \cite{ronneberger2015u} architecture with four down-sampling and four up-sampling layers, trained with an initial learning rate of $1 e^{-5}$ using the RMSprop optimizer for up to 30 epochs in a mini-batch size of 32. The loss function was the sum of the pixel-wise cross-entropy loss and Dice coefficient loss. The best model was identified by monitoring the best validation Dice score and used on the test dataset. For the multi-label classification, we used a pre-trained ResNet18 architecture trained on the ImageNet dataset. Instead of normalizing the input images with our training set's mean and standard deviation, we used the ImageNet mean and standard deviation to make them compatible with the pre-trained ResNet18 architecture. We used an initial learning rate of $1 e^{-5}$ with the Adam optimizer and trained up to 50 epochs in a mini-batch size of 32. The best model was identified by monitoring the best validation mAP score and used on the test set. We used AL to sample the best examples for annotation for up to six rounds and seven rounds for segmentation and classification tasks, respectively, starting with 200 samples and adding 100 examples (budget -- b) at each round.

\subsubsection{COVID-QU-Ex:} 
All the input images were loaded as RGB, with all channels having the same gray-scale value. To bring the pixel values within the range of 0 to 1, we normalized each image by dividing all the pixels by 255. Additionally, we normalized the images by subtracting the mean and dividing by the standard deviation, both with values of 0.5. The same hyperparameters were used for both VAAL and M-VAAL as those used in BraTS, and we also downsampled the original $256 \times 256$ images to $128 \times 128$ for both models. For the downstream multi-class classification task, we utilized a pre-trained ResNet18 architecture that was trained on the ImageNet dataset. We trained the model with an initial learning rate of $1 e^{-5}$ using the Adam optimizer and a mini-batch size of 32 for up to 50 epochs. The best model was determined based on the highest validation overall accuracy score, and we evaluated this model on the test set. We employed an AL method that involved sampling up to seven rounds using an AL budget of 100 samples. The initial budget for classification before starting active sampling was 100 samples.

We have released the GitHub repository for our source code implementation \footnote{https://github.com/Bidur-Khanal/MVAAL-medical-images \label{source_code}}. We implemented our method using the standard Pytorch 12.1.1 framework in Python 3.8 and trained in a single A100 GPU (40 GB).

\subsection{Benchmarks and Evaluation Metrics}

For our study, we employed two baseline control methods: random sampling and VAAL -- a state-of-the-art task agnostic learning method \cite{sinha2019variational} that doesn't use any auxiliary information while training. To quantitatively evaluate the segmentation, multi-label classification, and multi-class classification, we measured the Dice score, mean Average Precision (mAP), and overall accuracy, respectively.

\section{Results}
\subsubsection {Segmentation:}
\begin{figure}[t!]
\begin{center}
\includegraphics[width=0.7\columnwidth]{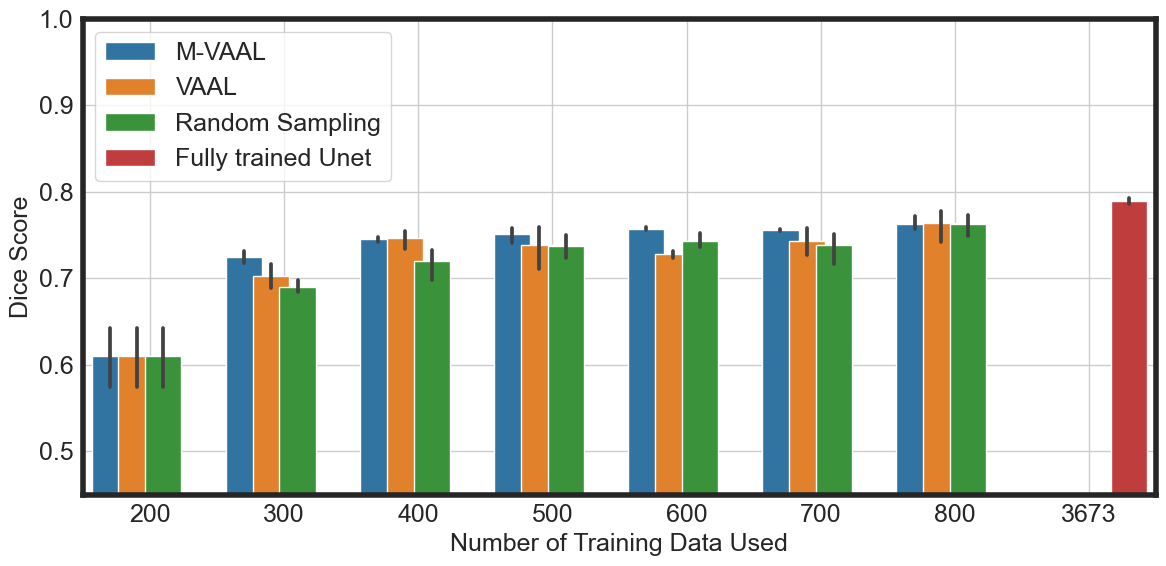}
\caption{{Whole tumor segmentation performance comparison of proposed (M-VAAL) method against the VAAL and random sampling baselines according to the Dice score.
\label{fig:segmentation_plot}%
}}
\end{center}
\end{figure}

Fig. \ref{fig:segmentation_plot} compares the whole tumor segmentation performance (in terms of Dice score) between our proposed method (M-VAAL), the two baselines (random sampling and VAAL), and a U-Net trained on the entire fully labeled dataset, serving as an upper bound with an average Dice score of $0.789$.

\begin{figure}[t!]
\begin{center}
\vspace{-0.5cm}
\includegraphics[width=0.9\columnwidth]{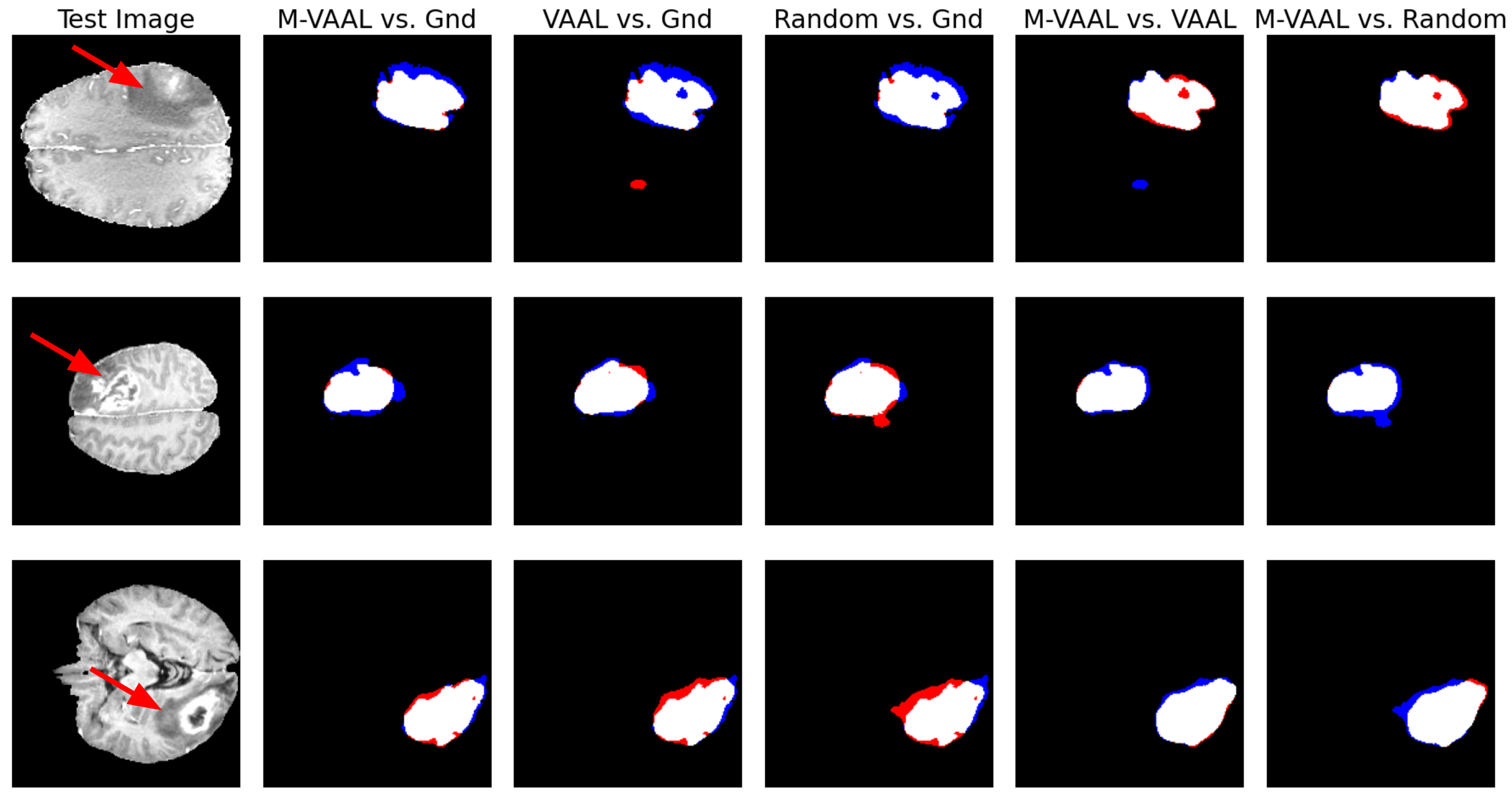}
\caption{{ Qualitative comparison of test segmentation masks generated by U-Nets trained on 400 samples using M-VAAL's selection method at the second round of AL phase. Columns 2-4 compare M-VAAl, VAAL, and random sampling against ground truth. Columns 5-6, compare M-VAAL against VAAL and random sampling. White denotes regions identified by both segmentation methods. Blue denotes regions missed by the test method (i.e., method listed first), but identified by the reference method (i.e., method listed second). Red denotes regions that were identified by the test method (i.e., method listed first), but not identified by the reference method (i.e., method listed second). Additionally, an arrow indicates the features that were segmented.
\label{fig:qualitative_results}%
}}
\end{center}
\end{figure}

As shown, with only $800$ labeled samples, the segmentation performance starts to saturate and approaches that of the fully trained U-Net on $100\%$ labeled data. Moreover, M-VAAL performs better than baselines in the early phase and gradually saturates as the number of training samples increase. 

Fig. \ref{fig:qualitative_results} illustrates a qualitative, visual assessment of the segmentation masks yielded by U-Net models trained with 400 samples selected by M-VAAL against ground truth, VAAL, and random sampling, at the AL second round. White denotes regions identified by both segmentation methods. Blue denotes regions missed by the test method (i.e., method listed first), but identified by the reference method (i.e., method listed second). Red denotes regions that were identified by the test method (i.e., method listed first), but not identified by the reference method (i.e., method listed second). As such, an optimal segmentation method will maximize the white regions, while minimizing the blue and red regions. Furthermore, Fig. \ref{fig:qualitative_results} clearly shows that the segmentation masks yielded by M-VAAL are  more consistent with the ground truth segmentation masks than those generated by VAAL or Random Sampling.

\subsubsection{Multi-label Classification:}
\begin{figure}[h!]
\begin{center}
\includegraphics[width=0.70\columnwidth]{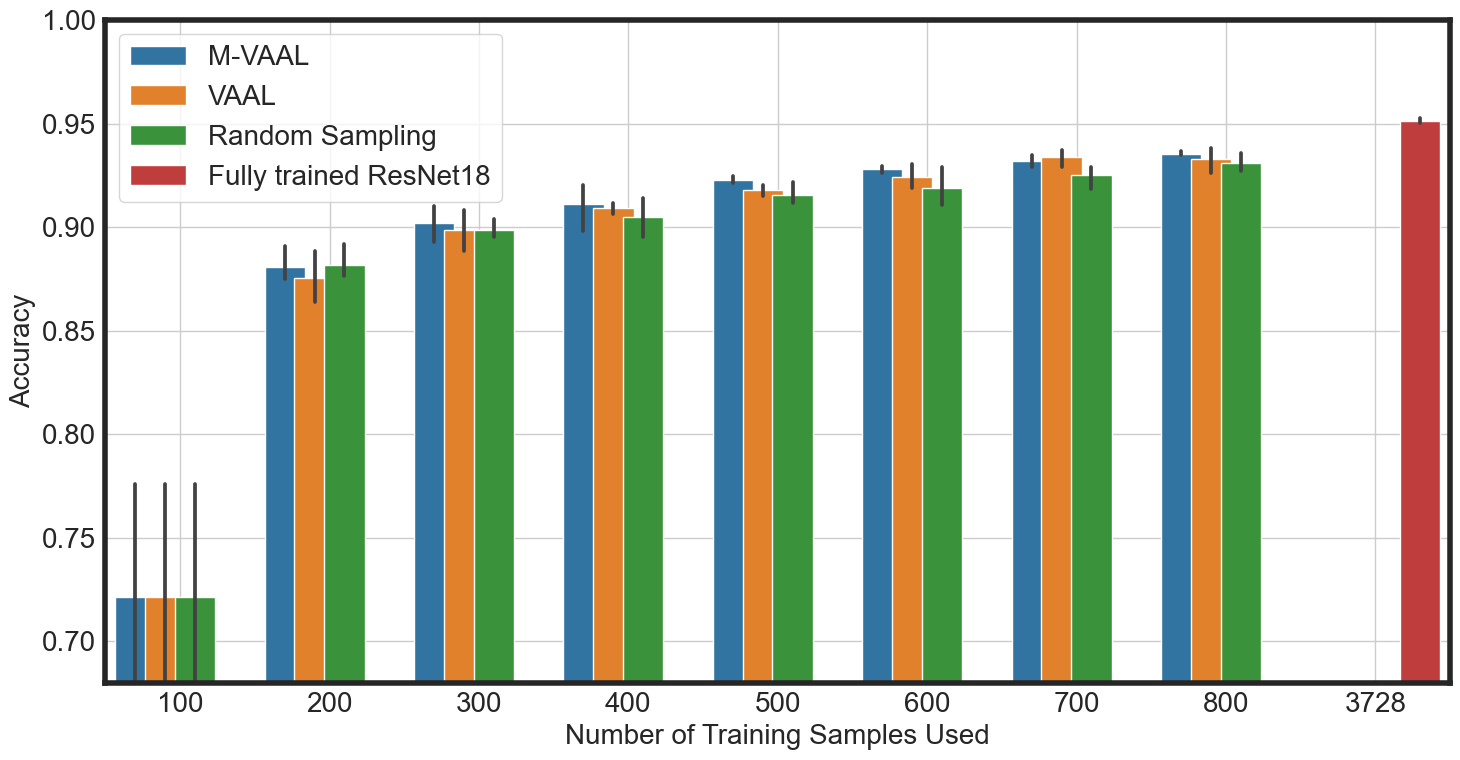}
\caption{{Comparison of the overall accuracy for multi-class classification of COVID, Non-COVID infections, and normal cases between the proposed M-VAAL method and two baseline methods (VAAL and random sampling).
\label{fig:classification_plot}%
}}
\end{center}
\end{figure}

In Fig. \ref{fig:classification_plot}, a comparison is presented of the multi-label classification performance, measured by mean average precision, between the M-VAAL framework and two baseline methods (VAAL and random sampling). The upper bound is represented by a fully fine-tuned ResNet18 network. It should be noted that, on average, M-VAAL performs better than the two baseline methods, particularly when fewer training data samples are used (i.e. 300 to 600 samples). 
The comparison models achieve a mean average precision close to that of a fully trained model even with a relatively small number of training samples, suggesting that not all training samples are equally informative. The maximum performance, which represents the upper bound, is achieved when using all samples and is $96.56\%$
\begin{figure}[h!]
\begin{center}

\includegraphics[width=0.7\columnwidth]{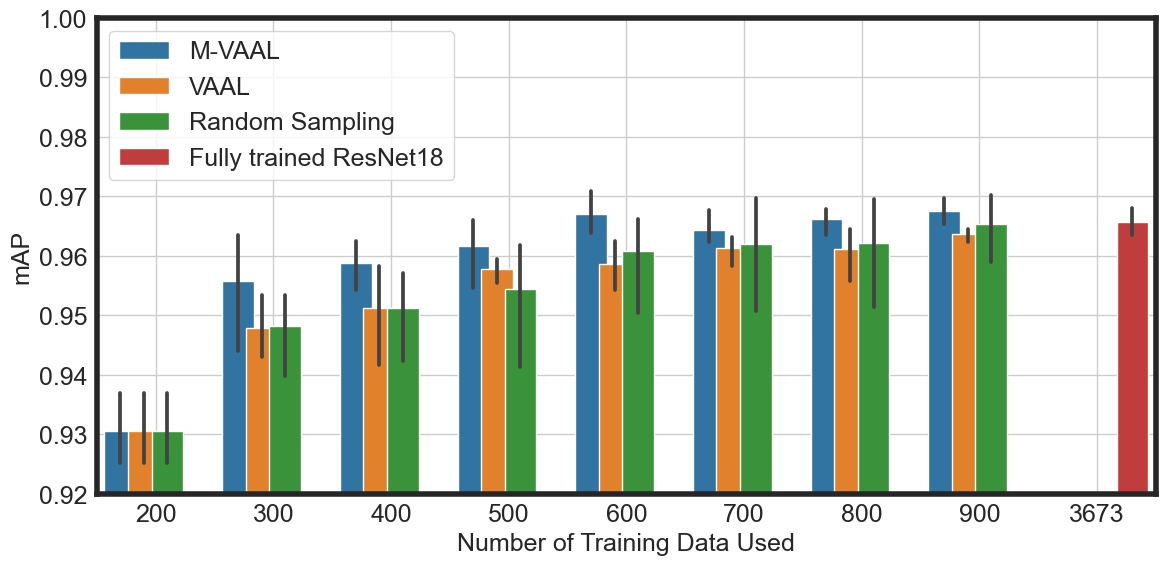}
\caption{{Comparison of the mean average precision (mAP) for multi-label classification of tumor types between the proposed M-VAAL method and two baseline methods (VAAL and random sampling).
\label{fig:classification_plot_covid}%
}}
\end{center}
\end{figure}
\subsubsection{Multi-class Classification:}
The classification performance of ResNet18 models in diagnosing a patient's condition using X-ray images selected by M-VAAL was compared with baselines. Fig. \ref{fig:classification_plot_covid} illustrates that M-VAAL consistently outperforms the other methods, albeit by a small margin. The maximum performance, which represents the upper bound, is achieved when using all samples and is $95.10\%$.

\section{Ablation Study: M-VAAL}
\label{ablation_study}

\begin{figure}[h!]
\centering
\begin{subfigure}[h!]{0.65\textwidth}
    \includegraphics[width=1\linewidth]{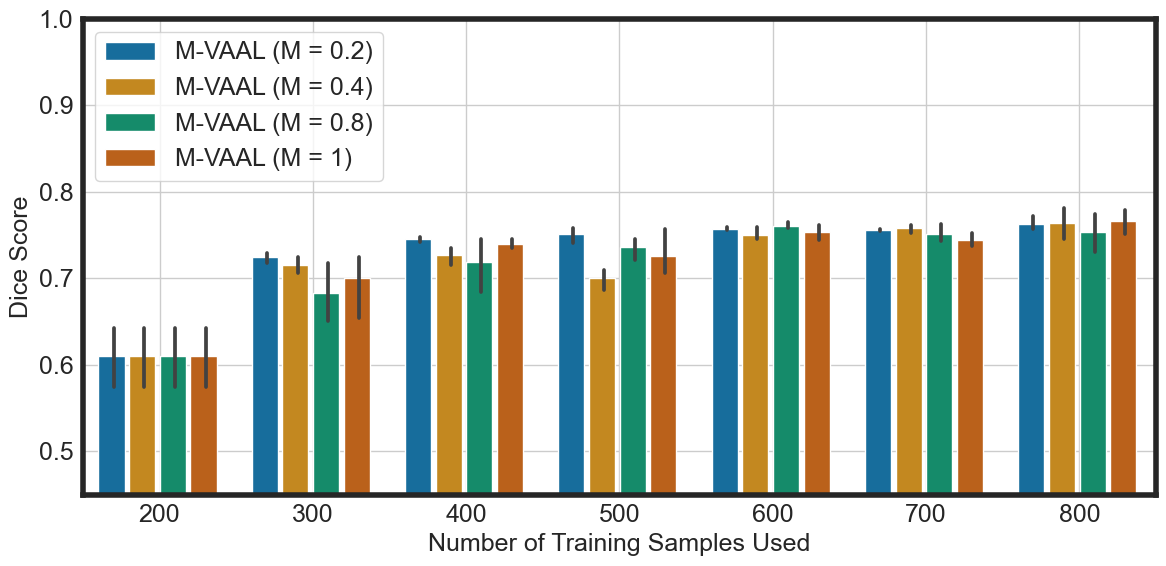}
    \label{fig:path_1}
    \vspace{-1.5em}
    \caption{}
  \end{subfigure}
\begin{subfigure}[h!]{0.65\textwidth}
    \includegraphics[width=1\linewidth]{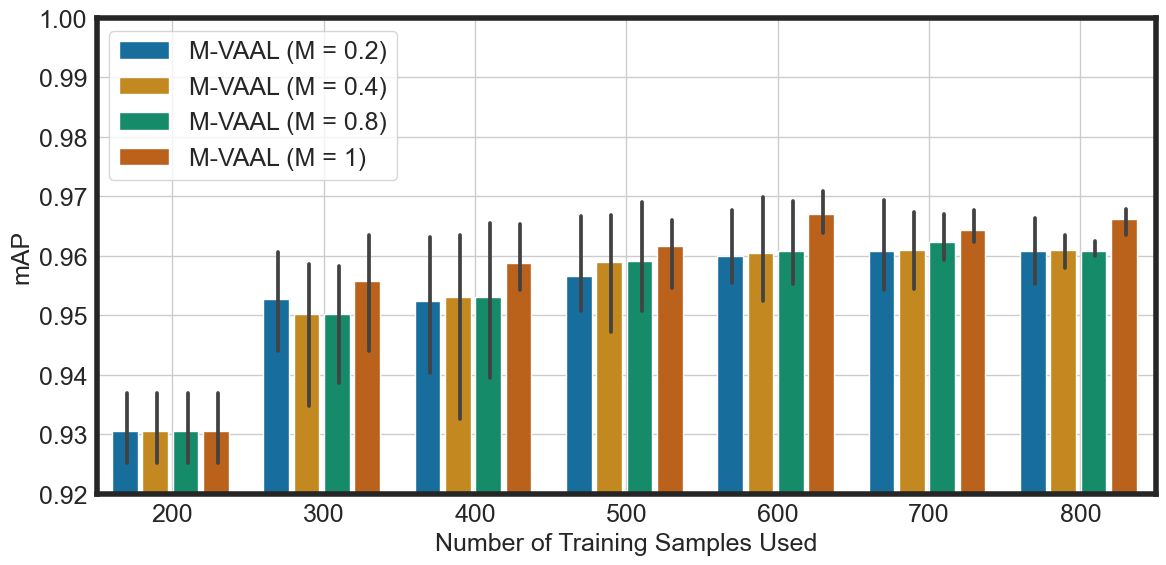}
    \label{fig:path_2}
    \vspace{-1.5em}
    \caption{}
  \end{subfigure}
  \begin{subfigure}[h!]{0.65\textwidth}
    \includegraphics[width=1\linewidth]{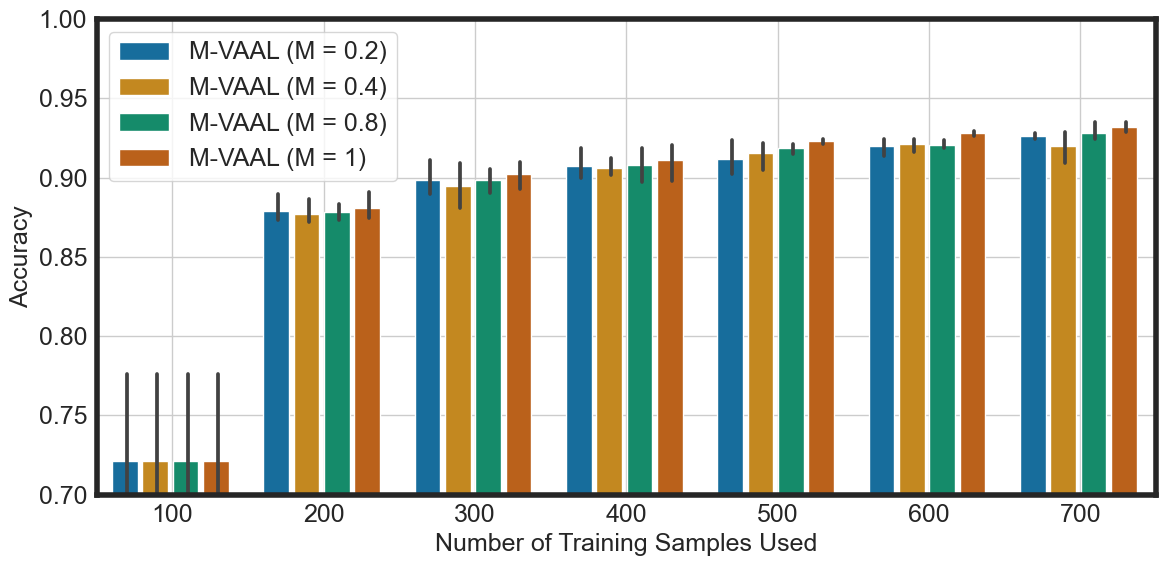}
    \label{fig:path_3}
    \vspace{-1.5em}
    \caption{}
  \end{subfigure}

\caption{Comparing the performance of (a) whole tumor segmentation, (b) tumor type multi-label classification, (c) chest X-ray infection multi-class classification with different training budgets, using the samples selected by M-VAAL trained with different values of M.}
\label{fig:albation_study}
\end{figure}

We conducted an ablation study to assess the effect of the multimodal loss component, which involved varying the hyperparameter $\gamma_3$ across a range of values (M = ${0.2,0.4,0.8,1.0}$). The results suggested that both VAAL and M-VAAL were sensitive to hyperparameters, which depended on the nature of the downstream task (see Fig. \ref{fig:albation_study}). For example, M = 0.2 performed optimally for the whole tumor segmentation task, while M = 1 was optimal for tumor-type multi-label classification and Chest X-ray image classification. We speculate that the multimodal loss components (see \ref{m1_decoder} and \ref{m2_decoder}) in the VAE loss serve as additional regularizers during the learning of latent representations. The adversary loss component, on the other hand, contributes more towards guiding the discriminator in sampling unlabelled samples that are most distinct from the distribution of the labeled dataset.

\section{ Discussion and Conclusion}
\label{discussion}

M-VAAL, while being sensitive to hyperparameters, samples more informative samples than the baselines. Our method is task-agnostic, but we observed that the optimal value of M differs for different tasks, as shown in Sec. \ref{ablation_study}. The type of dataset also plays an important role in the effectiveness of AL. While AL is usually most effective with large pools of unlabelled data that exhibit high diversity and uncertainty, the small dataset used in this study, coupled with the nature of the labels, led to high-performance variance across each random initialization of the task network. For instance, in the whole tumor segmentation task, as tumors do not have a specific shape, the prediction was based on texture and contrast information. Similarly, in the Chest X-ray image classification task, the images looked relatively similar, with only subtle minute features distinguishing between classes. In addition, the evaluation test also plays a crucial role in accessing the AL sampler. If the test distribution is biased and does not contain diverse test cases, the effective evaluation of AL will be undermined. We conducted several Student's t-tests to evaluate the statistical significance of the performance results; nevertheless, we observed high variance across runs in downstream tasks when smaller AL budgets are used as shown by the error bars in Fig. \ref{fig:classification_plot}, \ref{fig:segmentation_plot}, and \ref{fig:classification_plot_covid}, as a result, the scores are not always statistically significant, given the large variance.

In the future, we plan to investigate this further by evaluating the benefit of multimodal AL on a larger pool of unannotated medical data on diverse multimodal datasets that guarantee a diversified distribution. Additionally, we aim to explore the potential of replacing the discriminator with metric learning \cite{bellet2013survey} to contrast the labeled and unlabelled sets in the latent space. There is also a possibility of extending M-VAAL to other modalities. For instance, depth information can serve as auxiliary multimodal information to improve an AL sampler for image segmentation and label classification on surgical scenes with endoscopic images.

In this work, we proposed a task-agnostic sampling method in AL that can leverage multimodal image information. Our results on the BraTS2018 and COVID-QU-Ex datasets show initial promise in the direction of using multimodal information in AL. M-VAAL can consistently improve AL performance, but the hyperparameters need to be properly tuned.
\\

\noindent\textbf{Acknowledgements.} Research reported in this publication was supported by the National Institute of General Medical Sciences Award No. R35GM128877 of the National Institutes of Health, and the Office of Advanced Cyber Infrastructure Award No. 1808530 of the National Science Foundation. BB and DS are supported by the Wellcome/EPSRC Centre for Interventional and Surgical Sciences (WEISS) [203145Z/16/Z]; Engineering and Physical Sciences Research Council (EPSRC) [EP/P027938/1, EP/R004080/1, EP/P012841/1]; The Royal Academy of Engineering Chair in Emerging Technologies scheme; and the EndoMapper project by Horizon 2020 FET (GA 863146).

\bibliographystyle{splncs04}
\bibliography{mybib}


\newpage
\section{Supplementary Materials}

Here, we have compared the performance of M-VAAL with different baselines in three downstream tasks: brain tumor segmentation, multi-label tumor classification, and chest infection multi-class classification, in Table \ref{Table: seg}, Table \ref{Table: brast_multilabel}, and Table \ref{Table: covid_multiclass}, respectively. 
\begin{table}[h]
\begin{center}
\footnotesize
\caption{Summary of Dice Score (Mean $\pm$ Std. Dev.) for brain tumor segmentation between our proposed M-VAAL and baselines (VAAL and random sampling) at each active sampling round. }
\label{Table: seg}
\begin{tabular}{c|l|l|l}
\hline
& \multicolumn{3}{c}{Dice Score} \\\hline
 AL budget & M-VAAL        & VAAL          & Rand           \\\hline
$200$     & $0.611\pm0.028$ & $0.611\pm0.028$ & $0.611\pm0.028$     \\
$300$     & $\mathbf{0.725}\pm0.006$ & $0.702\pm0.011$ & $0.69\pm0.006$    \\
$400$     & $0.745\pm0.002$ & $\mathbf{0.747}\pm0.009$ & $0.72\pm0.016$   \\
$500$     & $\mathbf{0.751}\pm0.007$ & $0.739\pm0.02$  & $0.738\pm0.011$  \\
$600$     & $\mathbf{0.757}\pm0.001$ & $0.728\pm0.004$ & $0.743\pm0.007$   \\
$700$     & $\mathbf{0.756}\pm0.001$ & $0.743\pm0.013$ & $0.738\pm0.016$   \\
$800$     & $0.763\pm0.006$ & $\mathbf{0.764}\pm0.016$ & $0.763\pm0.011$  \\
\hline
\end{tabular}
\end{center}
\end{table}

\begin{table}[h]
\begin{center}
\footnotesize
\vspace{-1.0cm}
\caption{Summary of mAP (Mean $\pm$ Std. Dev.) for multi-label of tumor types between our proposed M-VAAL and baselines (VAAL and random sampling) at each active sampling round.}
\label{Table: brast_multilabel}
\begin{tabular}{c|l|l|l}
\hline
& \multicolumn{3}{c}{Mean Average Precision (mAP)} \\\hline
 AL budget & M-VAAL        & VAAL          & Rand          \\\hline
$200$     & $0.931\pm0.005$ & $0.931\pm0.005$ & $0.931\pm0.005$   \\
$300$     & $\mathbf{0.956}\pm0.008$ & $0.948\pm0.004$ & $0.948\pm0.006$    \\
$400$     & $\mathbf{0.959}\pm0.005$ & $0.951\pm0.007$ & $0.951\pm0.006$    \\
$500$     & $\mathbf{0.962}\pm0.005$ & $0.958\pm0.002$ & $0.954\pm0.009$   \\
$600$     & $\mathbf{0.967}\pm0.003$ & $0.959\pm0.003$ & $0.961\pm0.007$    \\
$700$     & $\mathbf{0.964}\pm0.002$ & $0.961\pm0.002$ & $0.962\pm0.008$    \\
$800$     & $\mathbf{0.966}\pm0.002$ & $0.961\pm0.004$ & $0.962\pm0.008$   \\
$900$     & $\mathbf{0.968}\pm0.002$ & $0.964\pm0.001$ & $0.965\pm0.005$   \\
\hline
\end{tabular}
\end{center}
\end{table}

\begin{table}[t!]
\begin{center}
\footnotesize
\vspace{-1.0cm}
\caption{Summary of Overall Accuracy (Mean $\pm$ Std. Dev.) for chest X-ray infection multi-class classification of tumor types between our proposed M-VAAL and baselines (VAAL and random sampling) at each active sampling round. }
\label{Table: covid_multiclass}
\begin{tabular}{c|l|l|l}
\hline
& \multicolumn{3}{c}{Overall Accuracy} \\\hline
 AL budget & M-VAAL        & VAAL          & Rand         \\\hline
$100$     & $0.722\pm0.049$ & $0.722\pm0.049$ & $0.722\pm0.049$ \\
$200$     & $0.881\pm0.007$ & $0.876\pm0.010$  & $\mathbf{0.882}\pm0.007$ \\
$300$     & $\mathbf{0.902}\pm0.007$ & $0.899\pm0.008$ & $0.899\pm0.004$   \\
$400$     & $\mathbf{0.911}\pm0.010$  & $0.909\pm0.002$ & $0.905\pm0.008$   \\
$500$     & $\mathbf{0.923}\pm0.001$ & $0.918\pm0.002$ & $0.916\pm0.005$   \\
$600$     & $\mathbf{0.928}\pm0.001$ & $0.924\pm0.006$ & $0.919\pm0.007$   \\
$700$     & $0.932\pm0.002$ & $\mathbf{0.934}\pm0.004$ & $0.925\pm0.005$   \\
$800$     & $\mathbf{0.935}\pm0.001$ & $0.933\pm0.005$ & $0.931\pm0.005$  \\
\hline
\end{tabular}
\end{center}
\end{table}


\end{document}